\documentclass[onecolumn, 11pt, journal]{IEEEtran}

\usepackage{amssymb}
\usepackage{amsmath}
\usepackage{epsfig}
\usepackage{url}

\begin{document}
\newtheorem{theorem}{Theorem}
\newtheorem{corollary}{Corollary}
\newtheorem{definition}{Definition}
\newtheorem{lemma}{Lemma}
\pagestyle{empty}

\def\QED{\mbox{\rule[0pt]{1.5ex}{1.5ex}}}
\def\proof{\noindent\hspace{2em}{\it Proof: }}

\date{}
\title{The Ergodic Capacity of Phase-Fading Interference Networks } 
\author{\normalsize  Syed A. Jafar\\
      {\small \it E-mail~:~syed@uci.edu} \\
       }
\maketitle
\thispagestyle{empty}

\begin{abstract}
We identify the role of equal strength interference links as bottlenecks on the ergodic sum capacity of a $K$ user phase-fading interference network, i.e., an interference network where the fading process is restricted primarily to independent and uniform phase variations while the channel magnitudes are held fixed across time. It is shown that even though there are $K(K-1)$ cross-links, only about $K/2$ disjoint and equal strength interference links suffice to determine the capacity of the network regardless of the strengths of the rest of the cross channels. This scenario is called a \emph{minimal bottleneck state}. It is shown that ergodic interference alignment is capacity optimal for a network in a minimal bottleneck state. The results are applied to large networks. It is shown that large networks are close to bottleneck states with a high probability, so that ergodic interference alignment is close to optimal for large networks. Limitations of the notion of bottleneck states are also highlighted for channels where both the phase and the magnitudes  vary with time. It is shown through an example that for these channels, joint coding across different bottleneck states makes  it possible to circumvent the capacity bottlenecks.

\end{abstract}
\section{Introduction}
The capacity of Gaussian interference networks is an active area of research in network information theory. Three vastly different perspectives have been used to approach this challenging problem. 
\begin{enumerate}
\item {\bf Elemental Networks:} The focus is on small networks - small number of users, finite SNR - in classical settings such as the two user interference network. This approach is most useful for exploring the microscopic details of optimal coding schemes. The goal is to obtain exact capacity characterizations, or capacity approximations within a constant number of bits. A representative recent result of this approach is the capacity characterization of the $2$ user interference network that is accurate to within $1$ bit  for all channel parameters \cite{Etkin_Tse_Wang}. A limitation of the elemental approach is that extensions to more general settings are not straightforward. For example, it has been shown that extensions to more than $2$ users or fading channels require fundamentally new concepts not encountered in the classical two user scenario.  Specifically, extensions to more than $2$ users involve the new concept of interference alignment \cite{Cadambe_Jafar_int}, while extensions to fading channels have to deal with the inseparability of parallel interference networks\cite{Cadambe_Jafar_inseparable}.
\item {\bf Degrees of Freedom:} The focus is on networks with arbitrary but finite number of users and asymptotically high signal and interference strengths relative to the noise floor at each receiver \cite{Jafar_Fakhereddin, Cadambe_Jafar_int}. This approach, essentially, offers insights into the scheduling/medium-access problem by identifying optimal ways of sharing signaling dimensions among competing users. The goal is to obtain capacity approximations within $o(\log(SNR))$, i.e., whose accuracy approaches 100\% as SNR approaches infinity.  A representative result of this approach is the degrees of freedom characterization of the $K$ user time-varying/frequency-selective interference network \cite{Cadambe_Jafar_int} 
\begin{eqnarray}
C_\Sigma&=&\frac{K}{2}\log(\mbox{SNR})+o(\log(\mbox{SNR}))
\end{eqnarray}
where $C_\Sigma$ is the sum-capacity of the network, $K$ is the number of users and $\mbox{SNR}$ is the signal to noise power ratio. A limitation of this approach is that while the degrees of freedom characterizations are dependent on the number of users through the capacity pre-log, they cannot be directly used to identify scaling laws with the number of users for any finite SNR. For finite SNR, the $o(\log(SNR))$ approximation error term can contain potentially arbitrary dependence on the number of users.
\item{\bf Network Scaling Laws:} The focus is on large networks, with the number of nodes approaching infinity \cite{Gupta_Kumar, Ozgur_Leveque_Tse}. This approach, essentially, offers fundamental insights into both scheduling (dense networks) and routing (extended networks) issues. The goal is to obtain approximations of the logarithm of the capacity within $o(\log(K))$. A recent representative result from this approach is the characterization
\begin{eqnarray}
\log(C_\Sigma)&=& \log(K) +o(\log(K))
\end{eqnarray}
for a dense network of $K$ users \cite{Ozgur_Leveque_Tse}. A limitation of this approach is that the dependence on SNR is  entirely lost in this perspective.
\end{enumerate}
In this work, we explore the exact ergodic capacity of an interference network, with arbitrary number of users and arbitrary SNR, albeit with the limiting assumption that the fading process is restricted primarily to phase variations while the channel magnitudes are held fixed. Combining the achievability results based on the recently proposed ergodic interference alignment scheme \cite{Nazer_Gastpar_Jafar_Vishwanath} with a converse argument based on a notion of bottleneck states introduced in this work, we find that the ergodic setting is surprisingly tractable and allows accurate and simple network capacity characterizations for broad classes of interference networks including many cases commonly considered. For large, dense networks it also provides a scaling law:
\begin{eqnarray}
\lim_{K\rightarrow\infty} \mbox{Prob}\left[\left|\frac{C_\Sigma}{K}-\frac{1}{2}\log(1+2\mbox{SNR})\right|>\epsilon\right]=0, ~~\forall \epsilon>0.
\end{eqnarray}
In other words,
\begin{eqnarray}
\frac{C_\Sigma}{K}\overset{\mathbb{P}}{\longrightarrow}\frac{1}{2}\log(1+2\mbox{SNR})
\end{eqnarray}
i.e., for large networks the capacity per user converges in probability to $\frac{1}{2}\log(1+2\mbox{SNR})$. Note that the dependence on the number of users $K$ as well as SNR is explicit here. For several networks of interest we show that this relationship represents the \emph{exact} non-asymptotic (i.e., arbitrary SNR, arbitrary $K$) capacity of the network as well.
\section{System Model}
\label{sec:systemmodel}
Consider the general $K$-user ergodic fading Gaussian interference network described by the input-output relationship:
\begin{eqnarray}
{Y}^{[r]}(n)&=&\sum_{t\in\mathcal{K}}H^{[rt]}(n) e^{j\phi^{[rt]}(n)}{ X}^{[t]}(n)+{ Z}^{[r]}(n),~~r\in\mathcal{K}, n\in\mathbb{N}\label{eq:systemmodel}
\end{eqnarray}
where at the $n^{th}$ channel use $ Y^{[r]}(n)$ and $Z^{[r]}(n)$ are the received symbol and additive white Gaussian noise (zero mean unit variance circularly symmetric complex Gaussian) seen by receiver $r$, $X^{[t]}(n)$ is the symbol transmitted from transmitter $t$, and $H^{[rt]}(n)$ and $\phi^{[rt]}(n)$ are the \emph{strength} and \emph{phase} of the channel between transmitter $t$ and receiver $r$, $t,r\in\mathcal{K}\triangleq \{1,2,\cdots,K\}$.  All symbols are complex. The power of the transmitted symbols is normalized to $1$, i.e., E$[|X^{[t]}|^2]\leq 1, \forall t\in \mathcal{K}$. The channel phase terms $\phi^{[rt]}(n)$ are i.i.d. and uniform over $[0,2\pi)$, and independent of the channel strengths. We make the following assumptions regarding the channel strengths:
\begin{eqnarray}
H^{[rt]}(n)&=&\sqrt{\mbox{INR}^{[rt]}(n)}, ~~r,t,\in\mathcal{K}, r\neq t\\
H^{[kk]}(n)&=&\sqrt{\mbox{SNR}}, ~\forall n\in\mathbb{N}, \forall k\in\mathcal{K}
\end{eqnarray}
The direct SNRs are held constant over time, primarily to avoid non-essential power-control issues. The normalization of SNRs so that all direct links have equal strength is somewhat restrictive but allows for exact capacity results without losing the essence of the problem.  It also alleviates fairness concerns always associated with sum capacity. Unless explicitly stated otherwise, no symmetry is assumed in general for the cross-channel strengths INR$^{[rt]}(n)$. Different cross-channel strengths can follow different distributions and may be correlated in space but are independent identically distributed (i.i.d.) in time.  Global channel knowledge is assumed at all transmitters and receivers.

There are $K$ independent messages $W_1, W_2, \cdots, W_K$. Message $W_k$ originates at transmitter $k$ and is intended for receiver $k$. The codewords are functions of the messages as well as the channel states. The probability of error, achievable rates $R^{[1]}, R^{[r]}, \cdots, R^{[K]}$, and sum capacity of the network $C_\Sigma$ are defined in the standard Shannon theoretic sense.

We define a \emph{channel state} as the matrix of channel strengths $H$ with elements $H^{[rt]}, ~r,t\in\mathcal{K}$.  We use the term ``user" to refer to a corresponding transmitter-receiver pair. We use the term capacity only in the sense of the sum capacity of the network. We refer to the fading coefficient $H^{[rt]}(n)e^{j\phi^{[rt]}(n)}$ between any transmitter-receiver pair as a channel or a link interchangeably. Channels $H^{[rt]}$ are called direct channels if $r=t$ and cross-channels if $r\neq t$. We say cross channels $H^{[rt]}(n)$ and $H^{[r't']}(n)$ are disjoint if and only if $\{r,t\}\cap\{r',t'\}=\{\}$, i.e., they do not involve a common user.
\section{Background}
We start by summarizing the background necessary for this work, which consists of three key ideas - the general idea of interference alignment, the inseparability of interference networks and a specific interference alignment scheme called ergodic interference alignment.
\subsection{Interference Alignment}
Evolving out of the degrees of freedom study of the $X$ channel \cite{MMK, Jafar_Shamai}, the idea of interference alignment was introduced in the context of the interference network in \cite{Cadambe_Jafar_int}. It refers to a design of signal vectors so that signals cast overlapping shadows at the receivers where they constitute interference while they remain distinguishable at the receivers where they are desired. The overlap between interferers at each receiver allows interference to be consolidated into a single entity at each receiver.  Since the desired signal at each receiver competes with only one effective consolidated interferer, it is possible for every user to access half the channel degrees of freedom --  i.e., \emph{everyone gets half the cake} \cite{Cadambe_Jafar_int}. 
\subsection{Inseparability of Interference Networks} For Gaussian point to point, multiple access and broadcast networks (with no common messages) ergodic capacity is the average of capacities over each channel state (subject to optimal power allocation across channel states). This implies that joint coding across channel states is not needed. However, for interference networks it was shown in \cite{Cadambe_Jafar_inseparable, Lalitha_inseparable} that the ergodic capacity is not the average of capacities over each channel state, i.e., parallel interference networks are in general not separable. The following example was provided in \cite{Cadambe_Jafar_inseparable}  to show that even a simple joint coding scheme where interference is treated as noise can achieve not only higher capacity but also more degrees of freedom than the most sophisticated separate coding scheme. Consider a three user interference network, with two parallel states given by channel matrices
\begin{eqnarray}
H=\sqrt{SNR}\left[\begin{array}{ccc}1&-1&1\\1&1&-1\\-1&1&1\end{array}\right]~~ H'=\sqrt{SNR}\left[\begin{array}{ccc}1&1&-1\\-1&1&1\\1&-1&1\end{array}\right]
\end{eqnarray}
It is shown in \cite{Cadambe_Jafar_inseparable} that separate coding for each channel state can only produce a sum capacity of $2\log(1+3\mbox{SNR})$, whereas the capacity with joint coding is $3\log(1+2\mbox{SNR})$. Further the joint coding capacity is achieved with the transmitters simply repeating the same symbol over the two sub-channels and the receivers adding their received signals from the two sub-channels. The key to the example is the complementary nature of the two channel matrices, i.e., $H+H'=2\sqrt{SNR}~I$ which simply aligns all the interference away from the desired signal. Note that while the signal power quadruples, the noise variance also doubles, so that the SNR doubles as a net effect.
\subsection{Ergodic Interference Alignment}
The pairing of complementary matrices exploited in the example above, is generalized to construct an ergodic interference alignment scheme in \cite{Nazer_Gastpar_Jafar_Vishwanath}. Essentially, for every channel matrix $H$ there is a complementary matrix $H'$,
\begin{eqnarray}
H=\left[\begin{array}{ccc}h_{11}&h_{12}&h_{13}\\h_{21}&h_{22}&h_{23}\\h_{31}&h_{32}&h_{33}\end{array}\right]~~ H'=\left[\begin{array}{ccc}h_{11}&-h_{12}&-h_{13}\\-h_{21}&h_{22}&-h_{23}\\-h_{31}&-h_{32}&h_{33}\end{array}\right]~~H+H'=2\left[\begin{array}{ccc}h_{11}&0&0\\0&h_{22}&0\\0&0&h_{33}\end{array}\right]
\end{eqnarray}
so that if the transmitters repeat the same symbol over the two channel states and the receivers add the outputs from the two channels then all interference is eliminated. With uniform phase distribution\footnote{The uniform distribution is not necessary, for pairing complementary states it suffices that the phase distribution is such that $p(\phi)=p(\phi+\pi)$ for all phase terms $\phi$.} the two states are equally likely. Channel state quantization enables a strong typicality argument that with high probability complementary pairs will occur in roughly equal numbers over a long codeword, thus making this alignment possible in an ergodic setting. Based on this approach, the following result is shown in \cite{Nazer_Gastpar_Jafar_Vishwanath}.

\begin{theorem}\cite{Nazer_Gastpar_Jafar_Vishwanath}\label{theorem:ach}
For the $K$ user interference network of (\ref{eq:systemmodel}), the following ergodic rates are achievable regardless of the cross-channel strengths:
\begin{eqnarray*}
R_k&=&\frac{1}{2}\log(1+2\mbox{SNR}),~~\forall k\in\mathcal{K}
\end{eqnarray*}
\end{theorem}
Theorem \ref{theorem:ach} states that regardless of the number of interferers or the strength of the interferers, each user is able to achieve the same rate that he would achieve if he had the channel to himself, with no interferers, \emph{half the time}. For $K=2$ users the achievability follows trivially by a channel state-independent TDMA scheme that allows each user to access the channel free from interference for half the time. However, for  $K>2$ users it is the ergodic interference alignment scheme that pairs complementary channel matrices, that is the key to achievability. Theorem \ref{theorem:ach} constitutes the achievability argument of all the capacity results in this paper. The main contributions of this paper are the converse arguments to show the optimality of ergodic interference alignment.

\section{Ergodic Capacity of Interference Networks}
\begin{figure}
\begin{minipage}[b]{0.5\linewidth} 
\centering
\includegraphics[width=6.5cm]{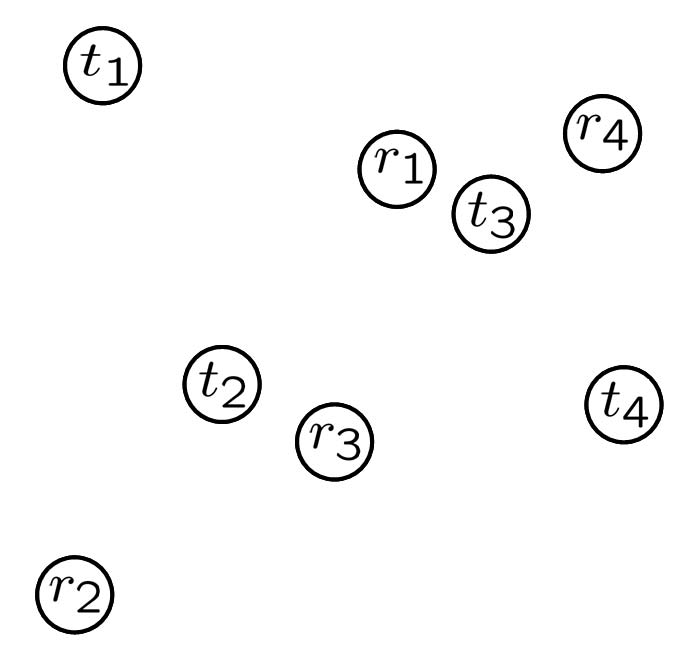}
\caption{4 User Interference Network. Distances indicate signal strengths.}\label{fig:botnet}
\end{minipage}
\hspace{0.5cm} 
\begin{minipage}[b]{0.5\linewidth} 
\centering
\includegraphics[width=5cm]{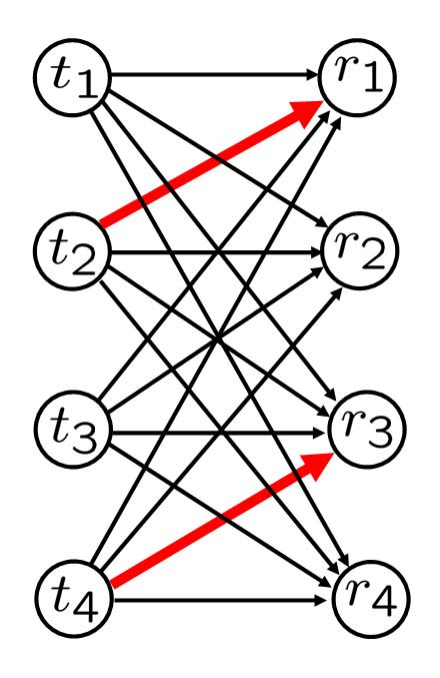}
\caption{Bottleneck state with bottleneck links highlighted}\label{fig:zbound}
\end{minipage}
\end{figure}

Ergodic interference alignment allows every user to achieve (slightly more than) half of their interference-free capacity at any SNR. However, the question remains whether this is the ergodic capacity of the interference network. Consider a $K=4$ user network example shown in Fig. \ref{fig:botnet} where the node distances may be translated into relative link strengths. From the figure, it is apparent that Receiver $2$ only sees weak interference while  receivers  $3$ and $4$ see both weak and strong interferers. It is known that strong interference can be decoded and cancelled, while weak interference can be treated as noise, in both cases without a significant rate penalty. A user experiencing very strong or very weak interference may be able to achieve close to his full interference-free capacity.  For example, in a network where all interferers are very strong, or all interferers are very weak compared to the desired signal strengths, the sum capacity is close to the sum of the users' interference-free capacities. It is clear therefore that achieving only half of the interference-free rate is not necessarily a sum-capacity optimal scheme.


\subsection{Bottlenecks}
For further insights into the optimality of ergodic interference alignment, consider the generalized degrees of freedom perspective (GDOF) introduced in \cite{Etkin_Tse_Wang} for the two user interference network and generalized in \cite{Jafar_Vishwanath_GDOF} to the symmetric $K$ user interference network. The capacity benefits of weak or strong interference are visible in the GDOF characterization. Indeed it is seen that  each user achieves strictly \emph{more} than half his interference-free degrees of freedom, except for two scenarios where $\alpha=\frac{\log(\mbox{\tiny INR})}{\log(\mbox{\tiny SNR})}$ takes values $1$ or $1/2$. In fact it is known that the network degrees of freedom ($\alpha$=1) are maximized by each user achieving half his interference-free degrees of freedom. This observation suggests that if the interference is of comparable strength to the desired signal then a scheme that achieves half the interference-free rate for each user may be close to optimal. This intuition forms the foundation for this work.  Indeed we establish the role of equal strength interferers as bottlenecks on the network capacity. The following definitions make this notion precise.
\begin{itemize}
\item{\bf Bottleneck Link:} The cross-channel $H^{[rt]}(n)$ between receiver $r$ and transmitter $t$, $t\neq r$, is a bottleneck link if INR$^{[rt]}(n)=$SNR, $\forall n\in\mathbb{N}$.
\item{\bf Bottleneck State:} The $K$-user interference network is in a bottleneck state if the network capacity does not depend on the distribution of the non-bottleneck link strengths. In particular, capacity is unchanged if we add more bottleneck links.
\item{\bf Reducible Bottleneck States:} A bottleneck state is reducible if relaxing a bottleneck link condition produces another bottleneck state. 
\item{\bf Minimal Bottleneck State:} A bottleneck state is minimal if there is no other bottleneck state with strictly fewer bottleneck links.
\end{itemize}
\begin{figure}[h]
\begin{center}
\includegraphics[width=4cm]{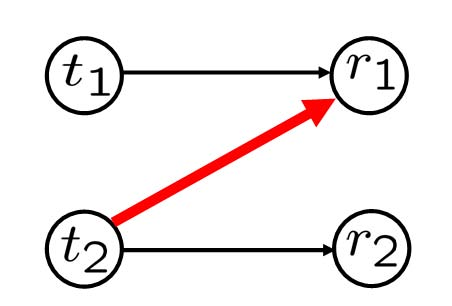}
\end{center}
\caption{Network of Lemma \ref{lemma:Z2} (only bottleneck and direct links are shown).}\label{fig:z2}
\end{figure}
\begin{lemma} \label{lemma:Z2}
The two user interference network with bottleneck link $H^{[12]}$,
\begin{eqnarray}
Y^{[1]}(n)&=&\sqrt{\mbox{SNR}}~e^{j\phi^{[11]}(n)}X^{[1]}(n)+\sqrt{\mbox{SNR}} ~e^{j\phi^{[12]}(n)}X^{[2]}(n)+Z^{[1]}(n)\\
Y^{[2]}(n)&=&\sqrt{\mbox{INR}^{[21]}(n)}~e^{j\phi^{[21]}(n)}X^{[1]}(n)+\sqrt{\mbox{SNR}}~e^{j\phi^{[22]}(n)}X^{[2]}(n)+Z^{[2]}(n)
\end{eqnarray}
has sum capacity
\begin{eqnarray*}
C_\Sigma&=&\log(1+2\mbox{SNR})
\end{eqnarray*}
regardless of the distribution of  $\mbox{INR}^{[21]}(n)$ values.
\end{lemma}
\proof
Achievability follows from Theorem \ref{theorem:ach} which is based on ergodic interference alignment. The converse is shown as follows. Consider any reliable coding scheme that can achieve arbitrary small probability of error by using appropriately long codewords. Let a genie provide receiver $2$ with user $1$'s message so he can eliminate all the interference due to $X^{[1]}(n)$. Since the coding scheme is reliable, receiver $1$ can also decode his own message and subtract the contribution from $X^{[1]}(n)$ from his received signal. This leaves the two receivers with statistically equivalent signals. Therefore, if receiver $2$ can decode message $W_2$ then so can receiver $1$. Since receiver $1$ is able to decode both messages, the sum rate of the coding scheme cannot be more than the sum rate capacity of the multiple access channel seen at receiver $1$, which is $\log(1+2\mbox{SNR})$.\hfill\QED

The implication of the term ``bottleneck" is clear from the definitions above - the bottleneck links determine the capacity of the network in a bottleneck state regardless of the strengths of the remaining cross-links. For example, in Lemma \ref{lemma:Z2} the link $H^{[12]}$ is a bottleneck link and the link strength of ${\bf H}^{[21]}$ is irrelevant. Note that we do require the phase of even the non-bottleneck links to vary according to a uniform distribution. This is not needed for the converse but it is needed for the achievability with ergodic interference alignment scheme. 

\begin{figure}
\begin{minipage}[b]{0.5\linewidth} 
\centering
\includegraphics[width=4cm]{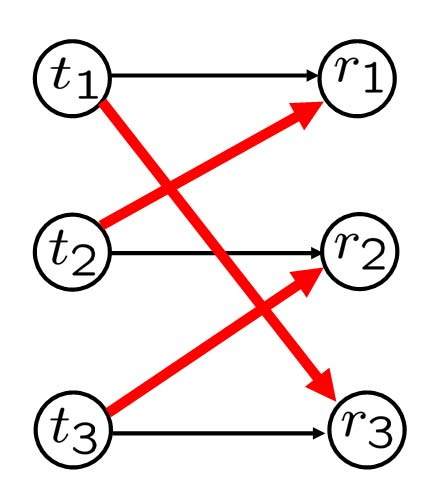}
\caption{Network of Lemma \ref{lemma:Z3} (bottleneck state) with only direct and bottleneck links shown.}\label{fig:z3}
\end{minipage}
\hspace{0.5cm} 
\begin{minipage}[b]{0.5\linewidth} 
\centering
\includegraphics[width=4cm]{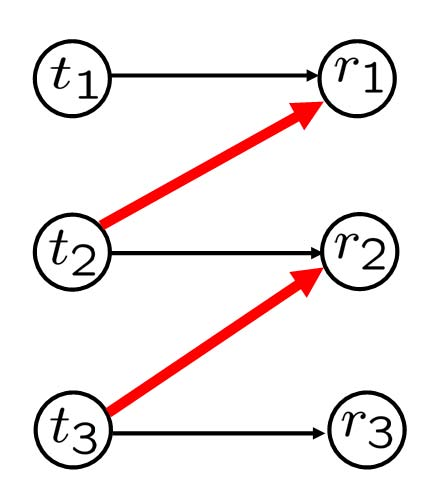}
\caption{Network of Lemma \ref{lemma:noZ3} (not a bottleneck state) with only direct and bottleneck links shown.}\label{fig:zn3}
\end{minipage}
\end{figure}

\begin{lemma} \label{lemma:Z3}
The three user interference network with three bottleneck links $H^{[12]}, H^{[23]}, H^{[31]}$  is in a bottleneck state
and has sum capacity
\begin{eqnarray*}
C_\Sigma&=&\frac{3}{2}\log(1+2\mbox{SNR})
\end{eqnarray*}
regardless of the distribution of the remaining $3$ cross-channel strengths.
\end{lemma}
The network is shown in Fig. \ref{fig:z3} with only direct and bottleneck links shown. Note that the non-bottleneck links not shown in the figure are not necessarily zero.

\proof
Achievability again follows from Theorem \ref{theorem:ach}. The converse is easily shown by considering two users at a time. For example, we bound the sum-rate of users $1,2$ by providing receivers $1,2$ with transmitter $3$'s message which allows receivers $1,2$ to remove interference from transmitter $3$ and leaves us with a two user interference network of the type considered in Lemma \ref{lemma:Z2}. This gives us the outer bound $R_1+R_2\leq \log(1+2\mbox{SNR})$. Similarly, outer bounds are obtained for the sum rates $R_2+R_3\leq\log(1+2\mbox{SNR})$ and $R_1+R_3\leq\log(1+2\mbox{SNR})$. Adding these three outer bounds gives us the sum rate capacity outer bound $C_\Sigma \leq \frac{3}{2}\log(1+2\mbox{SNR})$.\hfill\QED

\begin{lemma} \label{lemma:noZ3}
The three user interference network with two bottleneck links $H^{[12]}, H^{[23]}$ is not in a bottleneck state.
\end{lemma}

The network is shown in Fig. \ref{fig:zn3} with only direct and bottleneck links shown. Note that the non-bottleneck links not shown in the figure are not necessarily zero.

\proof
We prove the result by showing that different values of the non-bottleneck link strength $\mbox{INR}^{[31]}(n)$ produce different sum-capacity results, thus violating the definition of a bottleneck state. First, setting $\mbox{INR}^{[31]}=\mbox{SNR}, \forall n\in\mathbb{N}$ gives us the sum-capacity $C_\Sigma = \frac{3}{2}\log(1+2\mbox{SNR})$ by Lemma \ref{lemma:Z3}. Now alternatively, consider setting all non-bottleneck link strengths $\mbox{INR}^{[rt]}(n)=0, ~r\neq t, (r,t)\notin\{(1,2),(2,3)\}, \forall n\in\mathbb{N}$. In this channel let user $2$ not transmit at all. Then no interference is seen by receivers $1$ and $3$ allowing each of them to achieve their single user capacity. In other words, a sum-rate of $2\log(1+\mbox{SNR})$ is achievable. Since this achievable rate can be higher than $\frac{3}{2}\log(1+2\mbox{SNR})$, it is clear that the capacity of the $3$ user interference network of Lemma \ref{lemma:noZ3} depends on the strengths of the non-bottleneck links, i.e., it is not in a bottleneck state. \hfill\QED

Similarly, consider the $4$ user interference network example of Fig. \ref{fig:botnet}. As reflected by the node distances in Fig. \ref{fig:botnet} and highlighted explicitly in Fig. \ref{fig:zbound} there are two bottleneck links in this $4$ user network. By repeating the arguments of Lemma \ref{lemma:Z2} for each disjoint bottleneck link it is easily seen that the $4$ user network example of Fig. \ref{fig:botnet} is in a bottleneck state and its sum capacity is $2\log(1+2\mbox{SNR})$ regardless of the distribution of the non-bottleneck link strengths.

It is interesting to note that $2$ bottleneck links suffice for the $4$ user interference network but not for the $3$ user interference network, where a minimum of $3$ bottleneck links is required to put the network in a bottleneck state. This raises the question - what is the minimum number of bottleneck links that can put a $K$ user interference network into a bottleneck state? The bottleneck state with the minimum number of bottleneck links is called a minimal bottleneck state.  We characterize the minimal bottleneck state in the following theorem.

\begin{theorem} \label{theorem:minimal}
The minimal bottleneck states satisfy the following properties.
\begin{enumerate}
\item When $K$ is an even number, a $K$-user interference network in a minimal bottleneck state contains exactly $K/2$ bottleneck links and they are mutually disjoint. 
\item When $K$ is an odd number, a $K$-user interference network  in a minimal bottleneck state contains a total of  $(K+3)/2$ bottleneck links. 
\item Regardless of whether $K$ is even or odd, the ergodic sum capacity of a $K$-user interference network in a minimal bottleneck state is 
\begin{eqnarray*}
C_\Sigma&=& \frac{K}{2}\log(1+2\mbox{SNR})
\end{eqnarray*}
\end{enumerate}
\end{theorem}
{\it Remark: } Evidently, minimal bottleneck states represent classes of interference networks where ergodic interference alignment is exactly optimal.  Theorem \ref{theorem:minimal} shows that very few bottleneck links suffice to limit the capacity of a network.  For example, with $K=10$ users, a minimal bottleneck state needs only $5$ bottleneck links out of the total of $90$ cross-channels. If $5$ bottleneck links are disjoint, then the sum-capacity is determined regardless of the strengths of the remaining $85$ cross-channel coefficients. Moreover, there are $\frac{K!}{(K/2)!} = 30,240$ possible distinct minimal bottleneck states.

\vspace{1em}
\proof
Consider first the case that $K$ is even. We first show that there is a bottleneck state with $K/2$ disjoint links. Let the $K/2$ bottleneck links be $H^{[12]}, H^{[34]},H^{[56]},\cdots,H^{[K-1,K]}$. Now set the remaining links to zero, which cannot reduce the capacity. This gives us $K/2$ disjoint $Z$ interference networks, each with sum capacity $\log(1+2SNR)$ by Lemma \ref{lemma:Z2}. Thus the sum-capacity of this $K$ user interference network cannot be more than $\frac{K}{2}\log(1+2SNR)$ regardless of the strengths of the non-bottleneck links. However, we also know from Theorem \ref{theorem:ach} that the sum rate $\frac{K}{2}\log(1+2SNR)$ is achievable regardless of the strengths of the cross-links. Thus, the capacity of the $K$ user network with these $K/2$ bottleneck links is $\frac{K}{2}\log(1+2SNR)$ regardless of the strengths of the non-bottleneck links. It is therefore a bottleneck state.

To show that this is a minimal bottleneck state, we construct a proof by contradiction. Suppose there is a bottleneck state with $K/2-1$ bottleneck links. Then, since each bottleneck link can at most involve $2$ users, the total number of users involved can at most be $K-2$. Thus, there are at least 2 users that are not associated with any bottleneck link. Setting all non-bottleneck links to zero, we find that the sum-rate $\frac{K-2}{2}\log(1+2SNR)+2\log(1+SNR)$ is achievable. This is because the two users not involved with any interference can each achieve their interference-free capacity $\log(1+SNR)$ while the remainder $K-2$ users can achieve $\frac{1}{2}\log(1+2SNR)$ each by Theorem \ref{theorem:ach}. However, if we add a bottleneck link between the two remaining users, the sum capacity is only $\frac{K}{2}\log(1+2SNR)$. Since the capacity changes by adding a bottleneck link, this cannot be a bottleneck state. The contradiction completes the proof for the case that $K$ is even.

Next, consider the case that $K$ is odd. We first show that there is a bottleneck state with $(K+3)/2$ bottleneck links. Let us first designate $(K-3)/2$ bottleneck links $H^{[45]}, H^{[67]},\cdots,H^{[K-1,K]}$. This leaves three users who are assigned $3$ more bottleneck links $H^{[12]},H^{[23]}, H^{[31]}$ . Once again we set all non-bottleneck links to zero to find the outer bound on sum-capacity $\frac{K}{2}\log(1+2SNR)$. The outer bound for the sum-rate of users $4$ through $K$ follows from the disjoint two user $Z$ interference networks associated with each bottleneck link, each of which, from Lemma \ref{lemma:Z2}, has sum capacity $\log(1+2SNR)$. The outer bound on the sum-rate of users $1,2,3$ follows from Lemma \ref{lemma:Z3}. Once again by Theorem \ref{theorem:ach}, this rate is achievable regardless of the strengths of the non-bottleneck links. Thus, the sum-capacity of the network with a total of $(K-3)/2+3$ bottleneck links is fixed at $\frac{K}{2}\log(1+2SNR)$ regardless of the strengths of the non-bottleneck links. It is, therefore, a bottleneck state.

To show that this is a minimal bottleneck state, we construct a proof by contradiction. Suppose there is a bottleneck state with $\frac{K-3}{2}+2$ bottleneck links. Since $\frac{K-3}{2}$ bottleneck links can only involve at most $K-3$ users, we are left with $2$ bottleneck links that can be assigned and $3$ users. As shown by Lemma \ref{lemma:noZ3},  these $3$ users cannot be constrained by only two bottleneck links. The contradiction completes the proof.\hfill\QED

\subsection{Large Networks}
Clearly, bottleneck states are of academic interest because they allow an exact capacity characterization. However, the question remains whether these states play a significant role in a network where the link strengths are arbitrary. In particular, how many bottleneck links are there and more importantly, are there enough to put the network in a bottleneck state? Consider, for example an interference network with an even number of users. There are $\frac{K!}{(K/2)!}$ minimal bottleneck states. Is it enough to consider only minimal bottleneck states? Does every bottleneck state contain a minimal bottleneck state hidden behind some redundant bottleneck links ? The following theorem formally answers this question. 

 \begin{figure}[h]
\begin{center}
\includegraphics[width=4cm]{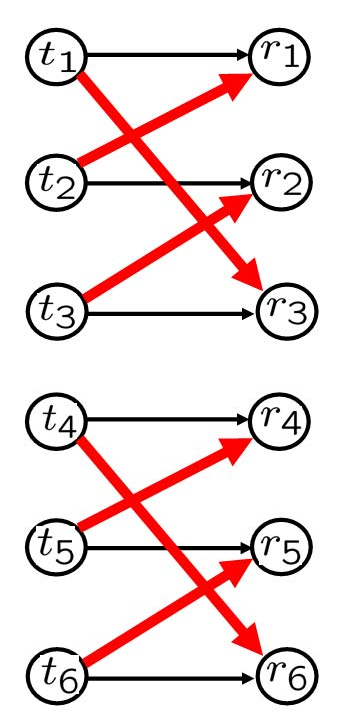}
\end{center}
\caption{Network of Theorem \ref{theorem:reduce} (only bottleneck and direct links are shown).}\label{fig:Z6}
\end{figure}

\begin{theorem} \label{theorem:reduce}
There exist irreducible bottleneck states that are not minimal bottleneck states.
\end{theorem}
{ Remark: In other words, while every minimal bottleneck state is irreducible, not every irreducible bottleneck state is a minimal  bottleneck state.}

\proof That a minimal bottleneck state is irreducible follows from definition. To show that an irreducible bottleneck state may not be a minimal bottleneck state, we offer an example. Consider the $6$ user interference network in a bottleneck state with $6$ bottleneck links as shown in Fig. \ref{fig:Z6}. From the result of Theorem \ref{theorem:minimal} we know that a minimal bottleneck state for a $6$ user interference network requires only $3$ bottleneck links. However, from the result of Lemma \ref{lemma:noZ3} it is easily seen that one cannot relax any of the $6$ bottleneck links in Fig. \ref{fig:Z6} to obtain another bottleneck state. \hfill\QED

Thus, there may be many more bottleneck states than minimal bottleneck states. Instead of counting the number of bottleneck states, we proceed to address directly the role of bottleneck states in determining the capacity of a network. Specifically we consider a \emph{dense network}, which we define to be a  wireless network with a large number of users where cross-channel strengths are determined i.i.d. according to some density function, but then they are held fixed for all time,
\begin{eqnarray}
\mbox{INR}^{[rt]}(n)=\mbox{INR}^{[rt]}, r,t \in\mathcal{K}, r\neq t, \forall n\in\mathbb{N}.
\end{eqnarray}
Note that the direct channel strengths are still all fixed at the same value SNR, and all phase variations are still i.i.d., uniform and independent of the channel strength. Because of the i.i.d. assumption on the density functions,  we call such a network a dense network. In addition, we make an important assumption, that for any constant $\eta >0$ and for all $r,t\in\mathcal{K}, r\neq t$,
\begin{eqnarray}
\mbox{Prob}(\mbox{INR}^{[rt]}\in(\mbox{SNR}-\eta, \mbox{SNR}+\eta))>0.\label{eq:support}
\end{eqnarray}

How likely is it that the dense network defined above will be in a bottleneck state? Requiring exact equality between INR and SNR makes the probability of a bottleneck link zero, so a network will never be exactly in a bottleneck state. A more useful question is to ask what is the probability that a network will be \emph{close} to a bottleneck state? Closeness here corresponds to the gap between the network capacity outer bound and the sum rate $\frac{K}{2}\log(1+2\mbox{SNR})$ achievable with ergodic interference alignment. Using a relaxed version of the bottleneck concept, the following result is obtained.
\begin{theorem} \label{theorem:dense}
For a dense $K$ user interference network, the sum-capacity per user $C_\Sigma/K$ converges in probability to $\frac{1}{2}\log(1+2\mbox{SNR})$.
\begin{eqnarray*}
\lim_{K\rightarrow\infty} \mbox{Prob}\left[\left|\frac{C_\Sigma}{K}-\frac{1}{2}\log(1+2\mbox{SNR})\right|>\epsilon\right]=0
\end{eqnarray*}
for every $\epsilon>0.$
\end{theorem}
Thus, ergodic interference alignment achieves the ergodic capacity of a large network in the limit as the number of users approaches infinity. Note that unlike most capacity scaling laws which describe only the exponent of $K$ in the large network limit and are completely insensitive to SNR, this scaling law establishes not only the scaling with number of users $K$  but also the precise SNR dependence of network capacity.

\vspace{1em}
\proof 
Since the achievability part is already established in \cite{Nazer_Gastpar_Jafar_Vishwanath}, we will focus only on the converse, i.e., the outer bound argument in this proof. Given the positive constant $\epsilon>0$ in the statement of the theorem, let us define arbitrary positive constants $\epsilon_1, \epsilon_2,\delta>0$ such that $\epsilon = \epsilon_1+\frac{1}{2}\epsilon_2$ and $\delta< \epsilon_2$. These definitions will be used in establishing the limiting behavior of various terms. It is important to note that these constants as well as SNR are held fixed while the number of users $K\rightarrow \infty$.

For the purpose of this theorem,  we relax the notion of a bottleneck state as follows. We say a cross-channel is a $\delta$-bottleneck link  if the sum capacity of the associated two user $Z$ interference network is within $\delta$ of $\log(1+2SNR)$, i.e., the sum capacity is less than $\log(1+2SNR)+\delta$.  Under the assumption about dense networks made in (\ref{eq:support}), the relaxed definition gives us a non-zero probability for a cross-link to be a $\delta$-bottleneck link. Let this probability be $\Delta>0$. Since the cross-links are generated i.i.d. each link has the same probability $\Delta$ of being an $\delta$-bottleneck. Each $\delta$-bottleneck $H^{[rt]}$ introduces the constraint:
\begin{eqnarray}
R^{[r]}+R^{[t]}&\leq& \log(1+2\mbox{SNR})+\delta\label{eq:bbound}
\end{eqnarray}
Adding these constraints we have:
\begin{eqnarray}
\sum_{r,t\in\mathcal{K}, r\neq t}I^{[rt]}(R^{[r]}+R^{[t]})&\leq& \left(\sum_{r,t\in\mathcal{K}, r\neq t}I^{[rt]}\right)(\log(1+2\mbox{SNR})+\delta)\label{eq:add}
\end{eqnarray}
Here $I^{[rt]}$ is an indicator function which takes value $1$ when channel $H^{[rt]}$ is a $\delta-$bottleneck link and $0$ otherwise. Clearly
\begin{eqnarray}
\mbox{E}[I^{[rt]}]=\Delta, ~~\forall r,t\in\mathcal{K}.\label{eq:Delta}
\end{eqnarray}
\allowdisplaybreaks
Let $$R_\Sigma\triangleq \sum_{k=1}^KR^{[k]}$$ be the sum-rate variable. From (\ref{eq:add}) we proceed as follows.
\begin{eqnarray}
\frac{R_\Sigma}{K}&\leq&\frac{R_\Sigma}{K}+\frac{\left(\sum_{r,t\in\mathcal{K}, r\neq t}I^{[rt]}\right)(\log(1+2\mbox{SNR})+\delta)}{2\Delta K(K-1)}-\frac{\sum_{r,t\in\mathcal{K}, r\neq t}I^{[rt]}(R^{[r]}+R^{[t]})}{2\Delta K(K-1)}\\
&=&\frac{1}{2\Delta K(K-1)}\sum_{k=1}^KR^{[k]}\left(2\Delta(K-1)-\sum_{t\in\mathcal{K}, t\neq k}I^{[kt]}-\sum_{r\in\mathcal{K}, r\neq k}I^{[rk]}\right)\nonumber\\
&&+\underbrace{\left(\frac{\sum_{r,t\in\mathcal{K}, r\neq t}I^{[rt]}}{\Delta K(K-1)}\right)}_\beta\left(\frac{1}{2}\log(1+2\mbox{SNR})+\delta\right)\\
&\leq&\frac{\log(1+\mbox{SNR})}{2\Delta K}\sum_{k=1}^K\left|\underbrace{2\delta-\frac{\sum_{t\in\mathcal{K}, t\neq k}I^{[kt]}-\sum_{r\in\mathcal{K}, r\neq k}I^{[rk]}}{K-1}}_{\alpha_k}\right|+\frac{\beta}{2}(\log(1+2\mbox{SNR})+\delta)\label{eq:usebound}\\
&=&\frac{\log(1+\mbox{SNR})}{2\Delta}\left(\frac{\sum_{k=1}^K\left|\alpha_k\right|}{K}\right)+\frac{\beta}{2}(\log(1+2\mbox{SNR})+\delta) \label{eq:keyeq}
\end{eqnarray}
In (\ref{eq:usebound}) we used the bound $R^{[k]}\leq\log(1+\mbox{SNR}), \forall k\in\mathcal{K}$. In the remainder of the proof we will show that, as $K\longrightarrow\infty$, the first term in (\ref{eq:keyeq}) is bounded above by $\epsilon_1$ and the second term is bounded above by $\frac{1}{2}(\log(1+2\mbox{SNR})+\epsilon_2) $ with probability approaching 1, thereby establishing the desired result that $\lim_{K\longrightarrow\infty}\mbox{Prob}(\frac{R_\Sigma}{K}\leq \frac{1}{2}\log(1+2\mbox{SNR}) + \epsilon)=1$.

Consider the term $\alpha_k$. Using (\ref{eq:Delta}), we find the mean and variance of $\alpha_k$ as follows.
\begin{eqnarray}
\mbox{E}[\alpha_k]=\mbox{E}\left[\frac{\sum_{t\in\mathcal{K}, t\neq k}I^{[kt]}+\sum_{r\in\mathcal{K}, r\neq k}I^{[rk]}}{K-1}-2\Delta\right]&=&0\\
\mbox{var}[\alpha_k]=\mbox{E}\left[\left(\frac{\sum_{t\in\mathcal{K}, t\neq k}I^{[kt]}+\sum_{r\in\mathcal{K}, r\neq k}I^{[rk]}}{K-1}-2\Delta\right)^2\right]&=&\frac{2\Delta(1-\Delta)}{K-1}
\end{eqnarray}
Applying Jensen's inequality in the form $\mbox{E}[|X|]\leq\sqrt{\mbox{E}[|X|^2]}$,
\begin{eqnarray}
\mbox{E}[|\alpha_k|]\leq\sqrt{\mbox{E}[|\alpha_k|^2]}&=&\sqrt{\frac{2\Delta(1-\Delta)}{K-1}}\\
\Rightarrow \mbox{E}\left[\frac{1}{K}\sum_{k=1}^{K}\left|\alpha_k\right|\right]=\mbox{E}[|\alpha_k|]&\leq&\sqrt{\frac{2\Delta(1-\Delta)}{K-1}}
\end{eqnarray}
The last inequality follows since $\alpha_k$ are identically distributed (they need not be independent).

The well known Markov's inequality states that for any positive random variable $X$ and an arbitrary constant $\mu>0$, the Prob$(X\geq \mu)\leq \frac{\mbox{E}[X]}{\mu}$. Next we apply this inequality.
\begin{eqnarray}
\mbox{Prob}\left(\frac{\log(1+\mbox{SNR})}{2\Delta}\left(\frac{\sum_{k=1}^K\left|\alpha_k\right|}{K}\right)\geq \epsilon_1\right)
&\leq& \frac{\log(1+\mbox{SNR})}{2\Delta\epsilon_1}\sqrt{\frac{2\Delta(1-\Delta)}{K-1}}
\end{eqnarray}
Equivalently
\begin{eqnarray}
\mbox{Prob}\left(\frac{\log(1+\mbox{SNR})}{2\Delta}\left(\frac{\sum_{k=1}^K\left|\alpha_k\right|}{K}\right)\leq \epsilon_1\right)
&\geq& 1-\frac{\log(1+\mbox{SNR})}{2\Delta\epsilon_1}\sqrt{\frac{2\Delta(1-\Delta)}{K-1}}\nonumber\\
&&\longrightarrow 1 \mbox{ as } K\rightarrow\infty\label{eq:term1}
\end{eqnarray}
Similarly, let us consider the term $\beta$.
\begin{eqnarray}
\mbox{E}\left[\beta\right]=\mbox{E}\left[{\left(\frac{\sum_{r,t\in\mathcal{K}, r\neq t}I^{[rt]}}{\Delta K(K-1)}\right)}\right]&=&1\\
\mbox{var}[\beta]=\frac{1-\Delta}{\Delta K(K-1)}
\end{eqnarray}
Chebyshev's inequality states the following. Let $X$ be a random variable with expected value $\mu$ and finite variance $\sigma^2$. Then for any real number $t > 0$,
\begin{eqnarray}
\mbox{Prob}(\left|X-\mu\right|\geq t)&\leq&\frac{\sigma^2}{t^2}
\end{eqnarray}
Applied to our setting, Chebyshev's inequality implies that for any positive constant $\epsilon_3$:
\begin{eqnarray}
\mbox{Prob}\left(\left|\beta-1\right|\geq \epsilon_3\right)&\leq&\frac{1-\Delta}{\Delta K(K-1)}\frac{1}{\epsilon_3^2}\\
\mbox{Prob}\left(\beta\geq 1+\epsilon_3\right)&\leq&\frac{1-\Delta}{\Delta K(K-1)}\frac{1}{\epsilon_3^2}\\
\mbox{Prob}\left(\beta\leq 1+\epsilon_3\right)&\geq&1-\frac{1-\Delta}{\Delta K(K-1)}\frac{1}{\epsilon_3^2}\\
&&\longrightarrow 1 \mbox{ as } K\rightarrow \infty
\end{eqnarray}
Substituting
\begin{eqnarray}
(1+\epsilon_3)&=&\frac{\log(1+2\mbox{SNR})+\epsilon_2}{\log(1+2\mbox{SNR})+\delta}\\
\Rightarrow \epsilon_3&=&\frac{\epsilon_2-\delta}{\log(1+2\mbox{SNR})+\delta}
\end{eqnarray}
and noting that $\epsilon_2>\delta$ by definition, we have that
\begin{eqnarray}
\mbox{Prob}\left(\frac{\beta}{2}(\log(1+2\mbox{SNR})+\delta)\leq \frac{1}{2}(\log(1+2\mbox{SNR})+\epsilon_2)\right)&\geq&1-\frac{1-\Delta}{\Delta K(K-1)}\left(\frac{\log(1+2\mbox{SNR})+\delta}{\epsilon_2-\delta}\right)^2\nonumber\\
&&\longrightarrow 1 \mbox{ as } K \longrightarrow \infty\label{eq:term2}
\end{eqnarray}

\noindent Combining (\ref{eq:keyeq}), (\ref{eq:term1}), (\ref{eq:term2}) we have,
\begin{eqnarray}
\lim_{K\rightarrow\infty}\mbox{Prob}\left(\frac{R_\Sigma}{K}\leq \epsilon_1+\frac{1}{2}(\log(1+2\mbox{SNR})+\epsilon_2)\right)=1\\
\Rightarrow \lim_{K\rightarrow\infty}\mbox{Prob}\left(\frac{R_\Sigma}{K}-\frac{1}{2}\log(1+2\mbox{SNR})\leq \epsilon\right)=1
\end{eqnarray}
where we used the definition $\epsilon  = \epsilon_1+\epsilon_2/2$.
Since this is true for all achievable rates, we have shown that
\begin{eqnarray}
\lim_{K\rightarrow\infty}\mbox{Prob}\left(\frac{C_\Sigma}{K}-\frac{1}{2}\log(1+2\mbox{SNR})\leq \epsilon\right)=1
\end{eqnarray}
which is the converse statement of Theorem \ref{theorem:dense}.\hfill\QED

{\it Remark:} An alternate proof of Theorem \ref{theorem:dense} is presented by Johnson, Aldridge and Piechocki in \cite{Piechocki_graph} based on the famous result of Erdos and Renyi  (Theorem 2 of \cite{Erdos_Renyi}) that a complete matching exists almost surely in a bi-partite graph where any two nodes independently have an edge present between them with a probability $\Delta$. Since a complete matching divides the network into disjoint $Z$ interference network pairs, each in a $\delta$-bottleneck state, it follows that the sum-capacity per user for the entire network is limited by sum-capacity per user of $Z$ interference network in a $\delta-$bottleneck state, which also establishes the converse part of Theorem \ref{theorem:dense}. In fact, as shown in \cite{Piechocki_graph} the proof based on complete matching states provides a faster rate of convergence than our proof outlined above. However, the approach used in our proof may be easier to generalize to a wider class of models, such as the node placement based model, as shown in \cite{Piechocki_general}.

\subsection{Time-Varying Signal Strengths}
The main goal of this section is to highlight a significant limitation of the concept of bottleneck states when applied to channels with time-varying magnitudes. In all cases considered so far, the time-variations for the direct links and the bottleneck states have been restricted just to phase variations.  (Note that non-bottleneck cross-channels may vary in both phase \emph{and magnitude} following any distribution). Constant signal strengths with random phases may be appropriate on wireless channels where the propagation is dominated by only one strong path (e.g. the line-of-sight path) whose strength stays relatively constant while its phase may vary significantly due to small scale mobility. When multiple paths are present, however, the phase variations must be associated with channel strength variations as these paths add constructively and destructively even with small mobility. In such scenarios, the time variations of the direct links can be compensated by power control (e.g. channel inversion) which may justify our assumption of constant direct channel SNRs. The same is not true for cross-channels. It is not possible to simultaneously control transmit power so that both the direct links and the cross-links maintain constant strengths. Thus, it is important to consider fading models where   not only the cross-link phase but also the cross-link \emph{strengths/magnitudes} vary with time.

 \begin{figure}[h]
\begin{center}
\includegraphics[width=5.5in]{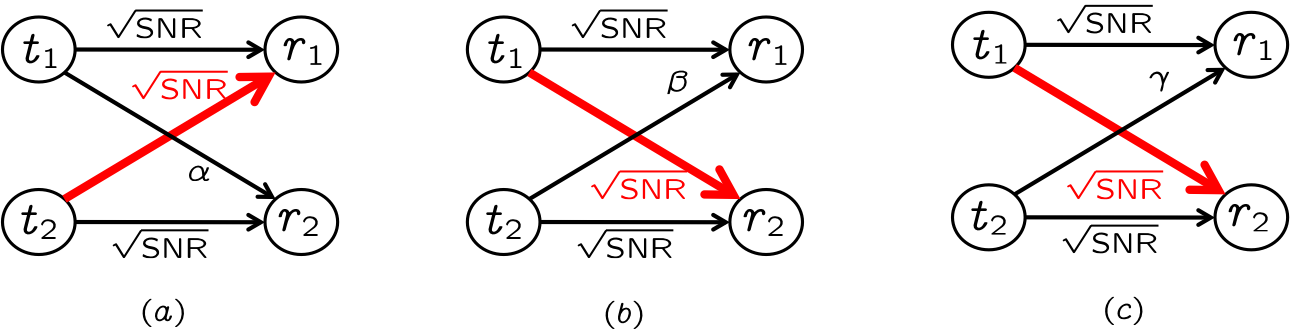}
\end{center}
\caption{Three different states of a  2 user interference network. While the capacity of each channel state by itself is limited by the bottleneck link, i.e., it does not depend on $\alpha, \beta, \gamma$, the joint capacity of the three states together is not limited by the bottleneck links, i.e., it depends also on $\alpha,\beta,\gamma$. Note that each channel coefficient also has a multiplicative random uniform time-varying phase term, omitted in the figure for compact notation.}\label{fig:bneckvary}
\end{figure}

From Theorem \ref{theorem:dense} we know that in a large (dense) network there are enough bottleneck links to put the network into a bottleneck state when the channel strengths are fixed. Now consider what happens when the channel strengths vary over time. While any fixed snapshot of the network state is likely to be in a bottleneck state, it turns out that this does not suffice to bound the capacity of the network whose state varies across the different bottleneck states. As an example consider Fig. \ref{fig:bneckvary} where we see  three different states of a $2$ user interference network  indicated by the following matrices. 
\begin{eqnarray}
H_a=\left[\begin{array}{cc}\sqrt{SNR}&\sqrt{SNR}\\\alpha&\sqrt{SNR}\end{array}\right],~~H_b=\left[\begin{array}{cc}\sqrt{SNR}&\beta\\\sqrt{SNR}&\sqrt{SNR}\end{array}\right],~~H_c=\left[\begin{array}{cc}\sqrt{SNR}&\gamma\\\sqrt{SNR}&\sqrt{SNR}\end{array}\right]
\end{eqnarray}
Note that while only channel magnitudes are highlighted above,  i.i.d. uniform phase fading is assumed for each channel coefficient within each state. For example, the signal at Receiver 1 in channel state $b$ is
\begin{eqnarray}
Y^{[1]}_b(n)&=&\sqrt{SNR}~e^{j\theta^{[11]}_b(n)}X^{[1]}_b(n)+\beta e^{j\theta^{[12]}_b(n)}X^{[2]}_b(n)+Z^{[1]}_b(n)
\end{eqnarray}
For this example, we assume that the channel changes state between these three states, so that each state is held for 1/3 of the time. 

As a baseline, first note that if the channel were to be held fixed permanently in any one of these three states, then because each state is a bottleneck state, the channel will have capacity $\log(1+2\mbox{SNR})$. In particular, note that the capacity of each channel state is limited by the bottleneck link so that it does not depend on the strengths, $\alpha, \beta, \gamma$, of the non-bottleneck cross-links. 

However, now consider the time-varying setting, where each of the three states $H_a, H_b, H_c$ is held for $1/3$ of the time. In this case, it turns out that the capacity is not limited by the bottleneck links, i.e., it depends on $\alpha,\beta,\gamma$ as well. To see this, consider the following distinct choices for $\alpha, \beta, \gamma$.
\begin{eqnarray}
\mbox{Choice 1: } (\alpha,\beta,\gamma)&=&(\sqrt{\mbox{SNR}}, \sqrt{\mbox{SNR}}, \sqrt{\mbox{SNR}})\\
\mbox{Choice 2: } (\log(1+\alpha^2),\beta,\gamma)&=&(2\log(1+\mbox{SNR}), 0, 0)
\end{eqnarray}
With the first choice, each state is identical and the capacity of the interference network is simply $\log(1+2\mbox{SNR})$. However, with the second choice, we argue that a sum rate of $\frac{4}{3}\log(1+\mbox{SNR})$ is achievable, which is \emph{higher than the capacity of any bottleneck state} at high SNR.  The scheme works as follows. Over state $H_a$ transmitter 1 sends his message $W_1$ to receiver $2$ at rate $2\log(1+SNR)$ which is less than the capacity of the cross-link between Transmitter 1 and Receiver $2$ for state $H_a$. Receiver $1$ and Transmitter $2$ are inactive in state $H_a$. Now, in states $H_b, H_c$ the users communicate with their desired receivers as they would in the absence of interference. Receiver $1$ does not see any interference and receiver $2$ having already decoded Transmitter $1$'s message is able to cancel the interference. Thus the sum-rate achieved is $4\log(1+\mbox{SNR})$ over three channel uses (one for each state). Equivalently, the rate achieved is  $\frac{4}{3}\log(1+\mbox{SNR})$ per channel use.

Thus, we have shown that the idea of bottleneck links, while very useful for the constant-magnitude varying-phase setting, does not generalize directly to varying-magnitude varying-phase  setting. While the bottleneck links, i.e., the equal strength interferers, limit the capacity when the channel magnitudes are held fixed, these limits may be circumvented by jointly encoding across different bottleneck states when the channel magnitudes are allowed to vary.

\section{Conclusion}
Ergodic interference alignment has recently been shown to achieve (slightly more than) half the interference-free capacity for an interference network at any SNR. We explore the optimality of ergodic interference alignment for ergodic capacity of an interference network where the channel coefficients undergo primarily phase-fading, i.e., the channel magnitudes are held fixed, while the phase terms are assume to vary according to an i.i.d. uniform distribution. The question arises because it is known that strong interference can be cancelled and weak interference can be treated as noise, without significant penalty to the users' rates. Thus, half of the interference-free rate is not necessarily optimal. The main insight offered in this paper is that network capacity is determined not by the weak or strong interferers but by the equal strength interferers. Thus, we identify equal strength interferers as the bottlenecks for network capacity. It is shown that very few bottleneck links suffice to determine the capacity of the network regardless of the strengths of the remaining links. This notion is formalized in the definition of a minimal bottleneck state. As an example, for a 10 user interference network, 5 disjoint bottleneck links determine the capacity regardless of the strengths of the remaining 85 interfering links. This capacity is achieved by ergodic interference alignment. We also consider a large network and show that it is always close to a bottleneck state, so that ergodic interference alignment is close to capacity optimal for large networks. 

The impact of time-variations in the \emph{magnitudes} of channel coefficients is explored next. The notion of bottleneck links is found to be insufficient to determine capacity bounds in this setting. The limitations of bottlenecks for channels with time-varying strengths are intrinsically related to the question of inseparability of parallel interference networks. It is shown in \cite{Cadambe_Jafar_inseparable, Lalitha_inseparable} that the capacity of time-varying interference networks is in general higher than the average of the capacity of the channel for each individual channel state, i.e., joint coding across channel states outperforms separate coding for each channel state. While the inseparability of bottleneck states is an important obstacle that prevents the generalization of the capacity bounds presented in this work to the case of time-varying channel magnitudes, there are some intriguing separability properties of bottleneck states highlighted in \cite{Jafar_ergodic} that may be useful to make further progress in this direction. 

\bibliographystyle{IEEEtran}

\begin{biographynophoto} {\bf Syed A. Jafar} (S  '99, M '04, SM '09) received the B. Tech. degree in Electrical Engineering from the Indian Institute of Technology (IIT), Delhi, India in 1997, the M.S. degree in Electrical Engineering from California Institute of Technology (Caltech) , Pasadena USA in 1999, and the Ph.D. degree in Electrical Engineering from Stanford University, Stanford, CA USA in 2003. His industry experience includes positions at Lucent Bell Labs , Qualcomm Inc. and Hughes Software Systems. He is currently an Associate Professor in the Department of Electrical Engineering and Computer Science at the University of California Irvine, Irvine, CA USA. His research interests include multiuser information theory and wireless communications.

Dr. Jafar received the NSF CAREER award in 2006, the ONR Young Investigator Award in 2008, the IEEE Information Theory Society paper award in 2009 and the Engineering School Fariborz Maseeh Outstanding Research Award in 2010. He received the UC Irvine Engineering Faculty of the Year award in 2006 and the UC Irvine EECS Professor of the Year Award twice, in 2009 and 2011, for excellence in teaching. Dr. Jafar was the inaugural instructor for the First Canadian School of Information Theory in 2011, a plenary speaker for the IEEE Communication Theory Workshop 2010 and SPCOM 2010, and a Visiting Erskine Fellow to the University of Canterbury New Zealand. He served as Associate Editor for IEEE Transactions on Communications 2004-2009, for IEEE Communications Letters 2008-2009 and is currently serving as Associate Editor for IEEE Transactions on Information Theory.
\end{biographynophoto}

\end{document}